\documentclass[lettersize,journal]{IEEEtran}
\usepackage{amsmath,amsfonts}
\usepackage{algorithmic}
\usepackage{algorithm}
\usepackage{array}
\usepackage{textcomp}
\usepackage{stfloats}
\usepackage{url}
\usepackage{verbatim}
\usepackage{graphicx}
\usepackage{cite}
\usepackage{subfigure}
\usepackage{hyperref}

\usepackage{booktabs} 
\usepackage{threeparttable} 

\usepackage{xcolor}

\newcommand{\markerone}{\textsuperscript{a}}
\newcommand{\markertwo}{\textsuperscript{b}}
\newcommand{\markerthree}{\textsuperscript{c}}
\newcommand{\markerfour}{\textsuperscript{d}}

\begin{document}

\title{Video-based Analysis Reveals Atypical Social Gaze\\ in People with Autism Spectrum Disorder}


\author{
    \IEEEauthorblockN{Xiangxu Yu\markerone*, Mindi Ruan\markertwo, Chuanbo Hu\markerthree, Wenqi Li\markertwo, Lynn K. Paul\markerfour, Xin Li\markerthree, and Shuo Wang\markerone}
    \\
    \IEEEauthorblockA{\markerone Department of Radiology, Washington University in St. Louis, St. Louis, 63110, MO, USA.}
    \\
    \IEEEauthorblockA{\markertwo Lane Department of Computer Science and Electrical Engineering, West Virginia University, Morgantown, 26506, WV, USA.}
    \\
    \IEEEauthorblockA{\markerthree Department of Computer Science, University at Albany, Albany, 12222, NY, USA.}
    \\
    \IEEEauthorblockA{\markerfour Humanities and Social Sciences, California Institute of Technology, Pasadena, 91125, CA, USA.}    
    \\
    \IEEEauthorblockA{*Corresponding author: xiangxu@wustl.edu}  
}

\markboth{}%
{Shell \MakeLowercase{\textit{et al.}}: A Sample Article Using IEEEtran.cls for IEEE Journals}

\IEEEpubid{}

\maketitle

\begin{abstract}
In this study, we present a quantitative and comprehensive analysis of social gaze in people with autism spectrum disorder (ASD). Diverging from traditional first-person camera perspectives based on eye-tracking technologies, this study utilizes a third-person perspective database from the Autism Diagnostic Observation Schedule, 2nd Edition (ADOS-2) interview videos, encompassing ASD participants and neurotypical individuals as a reference group. Employing computational models, we extracted and processed gaze-related features from the videos of both participants and examiners. The experimental samples were divided into three groups based on the presence of social gaze abnormalities and ASD diagnosis. This study quantitatively analyzed four gaze features: {\em gaze engagement, gaze variance, gaze density map, and gaze diversion frequency}. Furthermore, we developed a classifier trained on these features to identify gaze abnormalities in ASD participants. Together, we demonstrated the effectiveness of analyzing social gaze in people with ASD in naturalistic settings, showcasing the potential of third-person video perspectives in enhancing ASD diagnosis through gaze analysis.
\end{abstract}


\section{Introduction}

Autism spectrum disorder (ASD) is a complex neurodevelopmental disorder characterized by difficulties in social interaction, communication, and repetitive behaviors. The Autism Diagnostic Observation Schedule, Second Edition (ADOS-2), is a well-known assessment tool for ASD, considered a golden standard in clinical diagnosis. However, the manual administration of ADOS is time-consuming and requires specialized training, indicating the need for more efficient automated approaches. This research delves into utilizing computer science algorithms to assess gaze patterns in third-person videos from ADOS interviews automatically. By comparing these automated gaze analyses with ADOS scoring results, this study aims to pinpoint unique gaze characteristics in individuals with ASD, potentially improving diagnostic precision and effectiveness.

People with ASD demonstrate pervasive dysfunctions in social communication and interaction, restricted interests, and repetitive behaviors. To better understand the atypical social behavior in autism, there is an increasing trend to employ more natural and ecologically valid stimuli (e.g., complex scenes taken with a natural background) \cite{ames2010review} and to test participants in a more natural setting. On the one hand, using natural scene stimuli, people with ASD have demonstrated atypical gaze to social scenes \cite{birmingham2011comparing, chawarska2013decreased} and socially salient aspects of the scenes \cite{rice2012parsing, shic2011limited}, and reduced gaze towards threat-related scenes when presented with pairs of emotional or neutral images \cite{santos2012just}. Tasks presenting faces in a naturalistic setting demonstrate that people with ASD have reduced gaze to faces and the eye region of faces \cite{freeth2010gaze, klin2002visual, norbury2009eye, riby2013spontaneous, riby2009looking}. In particular, our prior study has provided a comprehensive investigation of eye tracking in autism \cite{wang2015atypical}: with over 5000 regions annotated in 700 images and a novel and sophisticated computational model, we have shown that compared with matched controls, people with ASD show a stronger image center bias regardless of object distribution, reduced saliency for faces and for locations indicated by social gaze, yet a general increase in pixel-level saliency at the expense of semantic-level saliency. Using dynamic natural scene videos, it has been shown that atypical gaze patterns in autistic adults are heterogeneous across but reliable within individuals \cite{keles2022atypical}.

On the other hand, recent studies have directly tested people with ASD during natural interactions with the experimenter. For example, when asking people with ASD to take photos in natural social communicative settings, they take abnormal photos of other people \cite{wang2016revealing}. Furthermore, there have been efforts to use video-based analysis of social behavior in children, where adult experimenter follows a semi-structured play interaction protocol designed to elicit a broad range of social behaviors in children \cite{rehg2013decoding} or during robot-assisted therapy sessions of children with autism \cite{marinoiu20183d}. Using simultaneous neuroimaging and eye-tracking with live in-person eye-to-eye contact, it has been shown that people with ASD have decreased right dorsal-parietal activity and increased right ventral temporal-parietal activity during live eye-to-eye contact \cite{hirsch2022neural}.

Manual assessment of ASD behaviors, such as assessing ADOS interviews, although extremely valuable, has various drawbacks, such as extensive training for assessors, scoring inconsistencies, and limitations in the number of assessments due to its labor-intensive nature. These challenges highlight the necessity for standardization and adaptability in diverse contexts, as evidenced by studies pointing out the need for consistent scoring metrics and adaptability in non-traditional settings \cite{mccrimmon2014test, gotham2009standardizing,schutte2015usability}. To meet this need for a more objective and quantitative assessment of ASD behaviors, several technologies have been applied. Research using eye-tracking technology has shown that people with ASD exhibit different gaze patterns, including decreased attention to individuals and faces, difficulty monitoring their gaze, and reduced focus on social signals \cite{Wang2017}. These findings highlight the value of eye-tracking technology in understanding specific visual attention processes in ASD \cite{falck2013eye, grynszpan2012self, wan2019applying}. However, employing specialized eye-tracking equipment for research presents its difficulties. These devices are expensive and restrict the range of experiments, often failing to capture genuine social interactions. Researchers have employed first-person perspective cameras and VR-based systems to study social gaze and interaction in individuals with ASD, revealing that such technologies can provide nuanced insights into the gaze behaviors and social communication of participants with ASD \cite{ahn2023objective, lahiri2011design,babu2017gaze,edmunds2017brief}. Moreover, while using first-person wearable cameras or face-to-face camera setups for studying interactions offers valuable insights, these methodologies introduce specific challenges, such as the need for particular experimental conditions and the potential influence of the camera's presence on participants' behavior, which might affect the genuineness of the observed interactions.

Although some impaired components of social communicative functioning have been reported, there is a lack of comprehensive analysis of social behavior in autism, especially in naturalistic settings. In this study, we analyzed videos from participants with ASD during real social interactions and employed sophisticated computational methods to analyze diverse social behaviors. Specifically, the current gold-standard diagnostic tool for clinical and research evaluations of ASD is the ADOS \cite{lord2009autism}, which involves standardized activities between the examiner and patient that take 40-60 minutes to complete \cite{lord1989austism}. ADOS professionals then rate the patient’s behaviors in various categories (e.g., social communication, repetitive behavior, speech, etc.) and provide a formal diagnosis of ASD based on the scores. Notably, ADOS can elicit and capture the most ASD-relevant behaviors.

Instead of utilizing eye trackers or first-person camera devices, we directly utilize objective measurement techniques to identify and extract gaze behaviors and characteristics from video recordings. In computer vision, there is extensive research on estimating gaze within videos. For example, Gaze360 presents a robust technique for estimating 3D gaze in diverse settings, utilizing a comprehensive dataset to analyze gaze patterns across different environments \cite{kellnhofer2019gaze360}. Similarly, LAEO-Net concentrates on recognizing mutual gaze in videos to comprehend social interactions, employing a deep CNN that considers spatial and temporal aspects \cite{marin2019laeo}. Furthermore, the Faze framework customizes gaze estimation through a few-shot learning strategy, adjusting to individual variations with minimal calibration and significantly improving model performance \cite{park2019few}. These studies underscore the progress in adaptive, inclusive, and context-aware gaze estimation solutions within computer vision.

In this study, we analyzed video recordings of ADOS-2 interviews from a third-person viewpoint, showing only the examiner and the participant with the examiner's face visible. Unlike previous eye-tracker-based approaches, our approach involves no additional specialized equipment other than standard video cameras. We utilized OpenFace \cite{baltruvsaitis2016openface, baltrusaitis2018openface}, a sophisticated toolkit for analyzing facial behavior, to identify gaze patterns in the video footage. OpenFace can accurately track gaze direction, eye coordinates, and other facial behavior indicators without special hardware. After extracting basic gaze data like gaze direction and eye coordinates, we applied our expertise in manual scoring to choose and compute various easily measurable hand-crafted features. These selected features were then utilized to delve deeper into the gaze behaviors of individuals with ASD and to assist in creating algorithms for predicting gaze, presenting a fresh approach to understanding and quantifying social attention during ASD evaluations. This method effectively decreases expenses and removes the necessity for extra specialized devices, thereby enhancing the accessibility and scalability of gaze estimation in a wide range of applications and environments.

We also utilized Google Speech-to-Text \cite{GoogleCloud2021} to extract conversations' timestamps from the simultaneous audio recordings, capturing moments when either the participant or examiner was speaking. Google Speech-to-Text is a robust tool that transforms spoken words into text, facilitating accurate speech synchronization with visual information. By merging these speech timestamps with the gaze attributes obtained from OpenFace, we could examine the variations in gaze behavior of individuals with ASD during speaking and non-speaking intervals. Furthermore, as a comparison group, we gathered video recordings of ADOS interviews with neurotypical volunteers. This enabled us to assess and compare the social gaze patterns between individuals with ASD and controls. By utilizing the gaze attributes derived from the recordings, we developed a Random Forest classifier to distinguish individuals with ASD from controls, demonstrating the potential of employing video-based gaze characteristics for classifying ASD.

In summary, we utilized video recordings of ADOS-2 interviews that depict interactions between individuals with autism and neurotypical individuals. We carefully identified, extracted, and examined the social gaze behaviors observed in these recordings by applying advanced gaze detection algorithms from computer vision. Our crucial analysis investigates the relationship between objective gaze measurement parameters and the corresponding ADOS-2 scores about gaze, emphasizing our contribution to improving the comprehension of gaze patterns in autism evaluations. This fusion of computer vision methods with clinical assessment represents a significant advancement in enhancing diagnostic instruments and approaches in autism research, demonstrating the potential of technological progress in supplementing traditional clinical evaluations.

\section{Results}

\begin{figure}
 \centering
 \subfigure[]{
   \label{DB:1}
   \includegraphics[width=0.48\columnwidth]{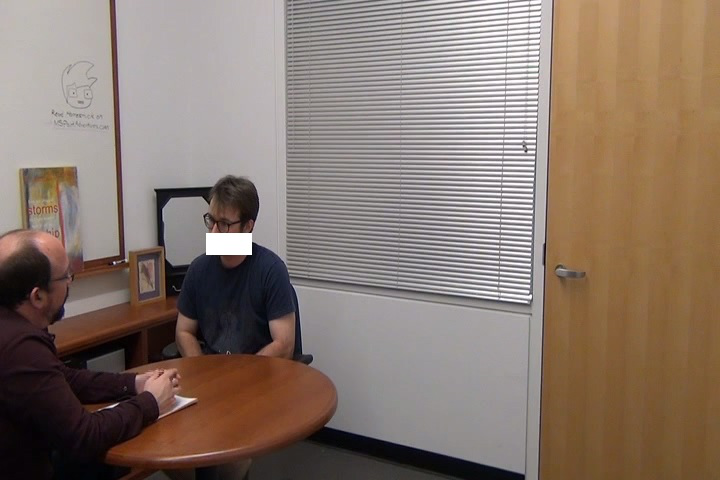}}
 \subfigure[]{
   \label{DB:2} 
   \includegraphics[width=0.48\columnwidth]{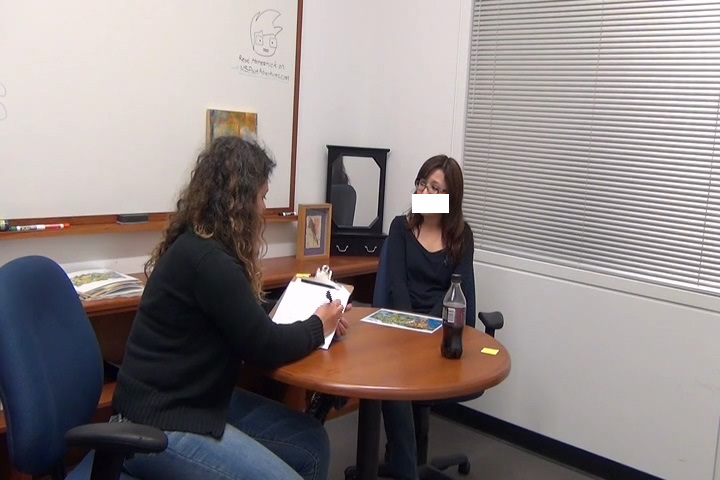}}
 \subfigure[]{
   \label{DB:3} 
   \includegraphics[width=0.48\columnwidth]{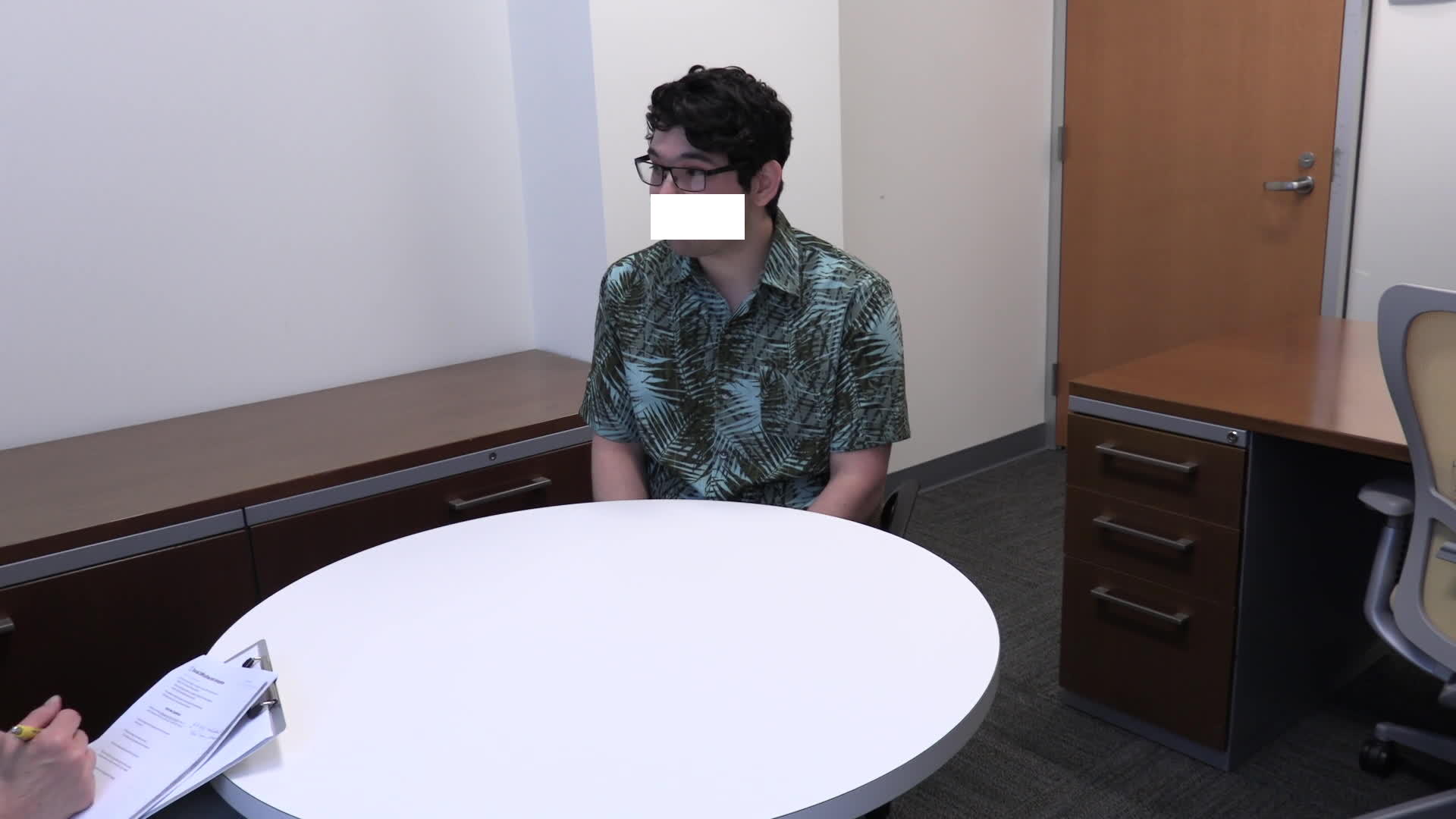}}
 \subfigure[]{
   \label{DB:4} 
   \includegraphics[width=0.48\columnwidth]{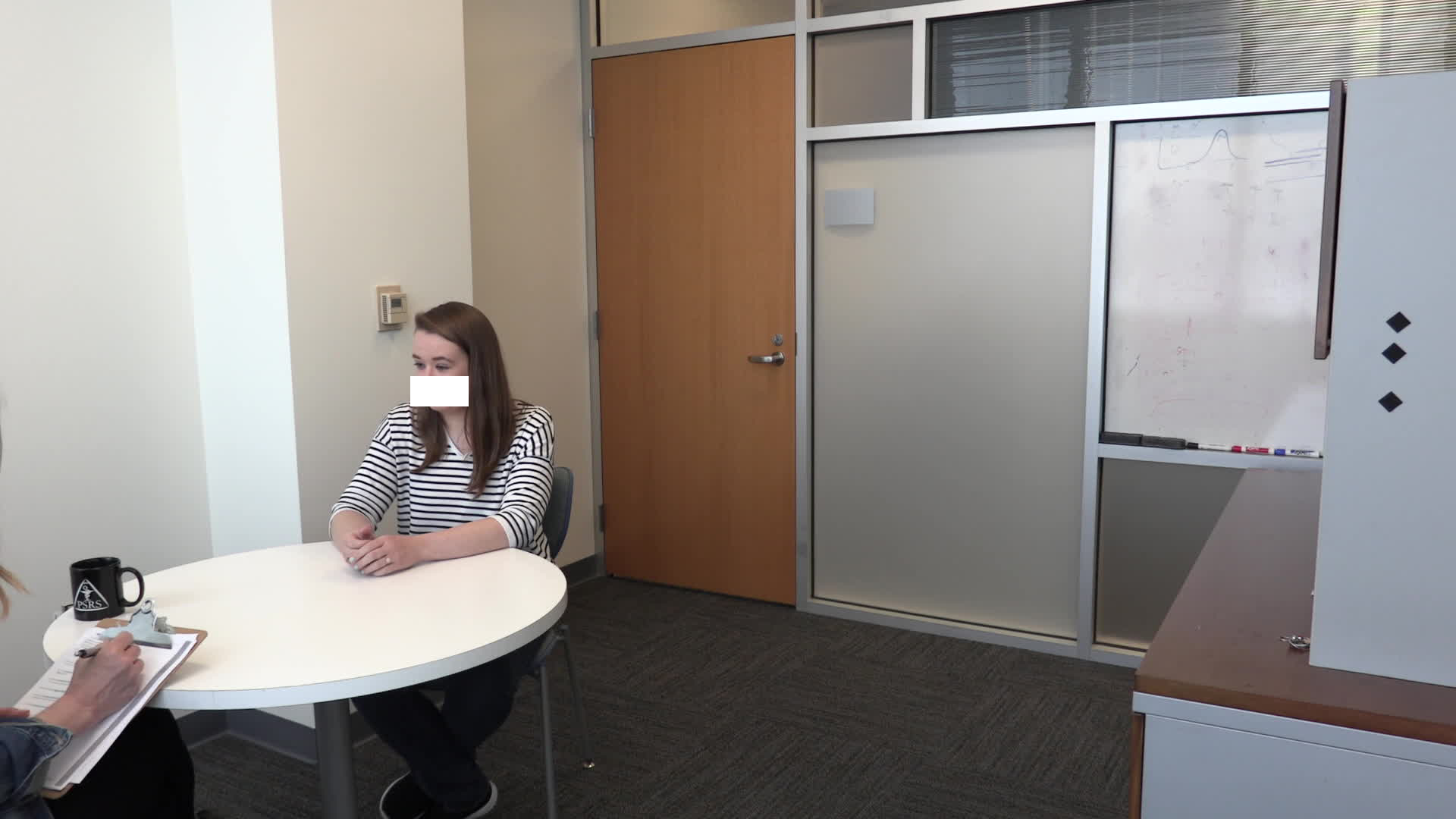}}
 \caption{Example video frames. (a, b) Caltech ADOS-2 Video Dataset. (c, d) WVU ADOS-2 Video Dataset.}
 \label{DB_example} 
\end{figure}

\begin{figure}
 \centering
 \subfigure[]{
   \label{raw:1}
   \includegraphics[width=0.48\columnwidth]{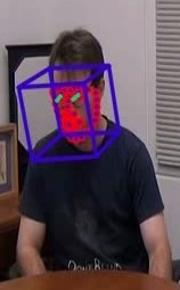}}
 \subfigure[]{
   \label{raw:2} 
   \includegraphics[width=0.48\columnwidth]{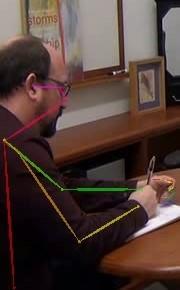}}
 \caption{Example frames of visualizing the raw multimodal features. (a) Facial/gaze features extracted using the OpenFace algorithm. (b) Action features extracted using the OpenPose algorithm.}
 \label{raw_example} 
\end{figure}

\begin{figure*}
\centering
\includegraphics[width = 1\textwidth]{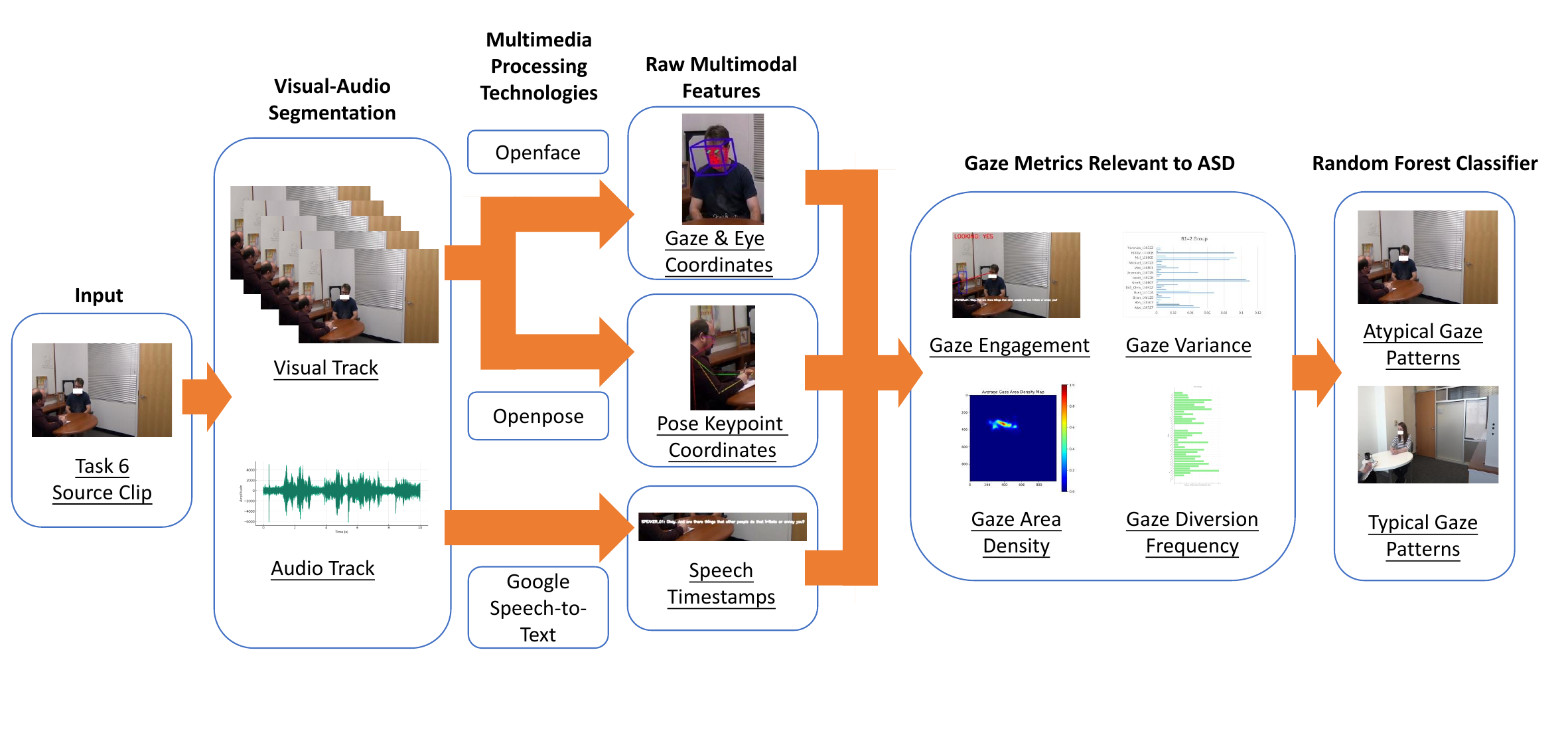}
\caption{Analysis pipeline. For each input video clip, we first separated the visual and audio tracks. The visual track, composed of all frames, and the audio track were further processed separately. Specifically, the visual track underwent feature detection using the OpenFace algorithm for facial/gaze features (Figure \ref{raw:1}) and the OpenPose algorithm for action features (Figure \ref{raw:2}), yielding raw multimodal features. The audio track was processed using Google's speech-to-text service for speech feature extraction. Subsequently, raw multimodal features were combined and processed to derive gaze metrics relevant to ASD based on domain knowledge of the condition. Ultimately, a random forest classifier was developed to differentiate between individuals exhibiting atypical and typical gaze patterns in the context of ASD.}
\label{gaze_flow}
\end{figure*}

\subsection{Overview}

In this study, we analyzed 43 samples of ADOS videos from individuals with ASD and 9 samples of ADOS videos from controls (see Section \ref{methods} for details). We focused on Task 6 (interview and discussion regarding "Social Difficulties and Annoyance"). The videos were divided into three groups based on the ADOS-2 scoring item B1, Unusual Eye Contact, score. The first group, known as the B1=2 group, contained 37 videos from Caltech participants who received a score of 2 on the B1 item, signifying atypical eye contact. The second group, the B1=0 group, encompassed 6 videos from Caltech participants with ASD who received a score of 0 on the B1 item, implying no obvious problems with eye contact. Finally, the control group included 9 videos from neurotypical participants supplied by WVU. This categorization facilitated a thorough comparison among groups, concentrating on the eye contact behavior of individuals with ASD in contrast to neurotypical controls.

Figure~\ref{DB:1} and \ref{DB:2} show example frames from the Caltech ADOS-2 Video Dataset and Figure ~\ref{DB:3} and Figure \ref{DB:4} show example frames from the WVU ADOS-2 Video Dataset. We used OpenFace (Figure \ref{raw:1}) to extract facial and gaze features (see Section \ref{methods} for gaze detection ratio showing reliable gaze detection) and OpenPose (Figure \ref{raw:2}) to extract action features. We also used Google Speech-to-Text to extract speech timestamps (see Section \ref{methods} for speaking duration ratio). These raw features were further processed and transformed into gaze metrics using a computational framework (see Figure \ref{gaze_flow} that illustrates the workflow for video processing for details). Below, we provide detailed characterizations of four aspects of social gaze and contrast the gaze characteristics of individuals with ASD with controls. A Random Forest classifier was further created and trained, with the ADOS-2 item B1 (Unusual Eye Contact) score as the predicted variable. The model's predictive results in discerning between individuals with ASD and the control group are presented. 

\begin{table*}[]
\caption{Mean and standard deviation of gaze metrics across the three groups.}
\label{results_table}
\centering
\begin{tabular}{ccccc}
\toprule
        & Gaze Engagement Ratio & Gaze Variance   & Gaze Concentration Area & Gaze Diversion Frequency \\ 
\midrule        
B1=2    & 41.29\%±31.20\%     & 0.0255±0.0333 & 5849.16±5771.93       & 3.0385±1.8270          \\
B1=0    & 67.67\%±30.44\%     & 0.0048±0.0031 & 3161.25±1442.79       & 3.3338±2.2311          \\
Control & 67.22\%±21.20\%     & 0.0103±0.0061 & 8530.02±5077.82       & 4.5065±2.0894          \\
\bottomrule
\end{tabular}
\end{table*}

\begin{table*}[]
\caption{Pairwise comparisons of gaze metrics across the three groups using Welch's t-test. ``df'' represents degrees of freedom, ``t'' is the t-statistic, ``p'' signifies the p-value, and ``d'' denotes the Cohen's d value. \textbf{Bold} indicates $p < 0.05$.}
\label{results_t-test}
\centering
\resizebox{\linewidth}{!}{
\begin{tabular}{ccccccccccccccccc}
\toprule
                & \multicolumn{4}{c}{Gaze Engagement Ratio} & \multicolumn{4}{c}{Gaze Variance} & \multicolumn{4}{c}{Gaze Concentration Area} & \multicolumn{4}{c}{Gaze Diversion Frequency} \\
\midrule                
                & df        & \textit{t}    & \textit{p}  & \textit{d}   & df     & \textit{t}  & \textit{p}   & \textit{d}  & df        & \textit{t}     & \textit{p}  &  \textit{d}  & df         & \textit{t}     & \textit{p}   &  \textit{d}  \\
\midrule                
B1=2 vs. B1=0    & 6.82      & -1.9619       & 0.0916   &  -0.8478   & 39.28  & 3.6754      & \textbf{0.0007}  & 0.6624  & 33.39     & 2.4067         & \textbf{0.0218}     &  0.4948  & 6.14       & -0.3078        & 0.7684     &  -0.1570  \\
B1=2 vs. Control & 17.56     & -2.9691       & \textbf{0.0084}  &  -0.8748    & 42.94  & 2.6075      & \textbf{0.0124}   & 0.5037  & 13.52     & -1.3816        & 0.1895     & -0.4743   & 11.17      & -1.9354        & 0.0787     &  -0.7819   \\
B1=0 vs. Control & 8.21      & 0.0314        & 0.9757  &  0.0179    & 12.53  & -2.2594     & \textbf{0.0423}  & -1.0535  & 9.82      & -2.9957        & \textbf{0.0137}     &  -1.3150  & 10.35      & -1.0228        & 0.3297     & -0.5467   \\
\bottomrule
\end{tabular}
}
\end{table*}

\begin{table*}[]
\caption{Mean and standard deviation of gaze metrics across the three groups, both while speaking and not speaking (note that gaze diversion frequency is not applicable for this analysis).}
\label{results_table_speaking_or_not}
\centering
\begin{tabular}{cclcclcclc}
\toprule
        & \multicolumn{3}{c}{Gaze Engagement Ratio}                       & \multicolumn{3}{c}{Gaze Variance}                         & \multicolumn{3}{c}{Gaze Concentration Area}                     \\
\midrule         
        & \multicolumn{2}{c}{Speaking}          & Non-Speaking            & \multicolumn{2}{c}{Speaking}        & Non-Speaking        & \multicolumn{2}{c}{Speaking}          & Non-Speaking            \\
\midrule           
B1=2    & \multicolumn{2}{c}{43.19\%±31.46\%} & 38.09\%±31.83\%       & \multicolumn{2}{c}{0.0233±0.0313} & 0.0282±0.0360     & \multicolumn{2}{c}{5540.00±4801.63} & 4148.62±3697.84       \\
B1=0    & \multicolumn{2}{c}{68.97\%±30.23\%} & 64.41\%±31.83\%       & \multicolumn{2}{c}{0.0045±0.0028} & 0.0058±0.0053     & \multicolumn{2}{c}{3164.09±1247.11} & 2625.77±1254.05       \\
Control & \multicolumn{2}{c}{67.55\%±20.90\%} & 64.95\%±22.54\%       & \multicolumn{2}{c}{0.0102±0.0006} & 0.0105±0.0075     & \multicolumn{2}{c}{8355.07±5136.85} & 6783.71±3110.01       \\
\bottomrule
\end{tabular}
\end{table*}

\begin{table*}[]
\caption{Pairwise comparisons of gaze metrics across the three groups while speaking and not speaking using Welch's t-test (note that gaze diversion frequency is not applicable for this analysis). ``df'' represents degrees of freedom, ``t'' is the t-statistic, ``p'' signifies the p-value, and ``d'' denotes the Cohen's d value. \textbf{Bold} indicates $p < 0.05$.}
\label{t-test_speaking_not_speaking}
\centering
\begin{tabular}{ccccccccccccc}
\toprule
\multicolumn{13}{c}{Speaking Interval} \\
\midrule
                & \multicolumn{4}{c}{Gaze Engagement Ratio} & \multicolumn{4}{c}{Gaze Variance}    & \multicolumn{4}{c}{Gaze Concentration Area} \\
\midrule                
                & df      & \textit{t}   & \textit{p}  & \textit{d}     & df    & \textit{t} & \textit{p}  &  \textit{d}  & df       & \textit{t}   & \textit{p}    &  \textit{d}  \\
\midrule
B1=2 vs B1=0    & 34.12   & -1.9265      & 0.0961  & -0.8232   & 38.99 & 3.5627     & \textbf{0.0009}  & 0.6401 & 32.14    & 2.5294       & \textbf{0.0165}  & 0.5256 \\
B1=2 vs Control & 25.96   & -2.8077      & \textbf{0.0116} & -0.8168  & 43.26 & 2.3652     & \textbf{0.0226}  & 0.4594 & 11.64    & -1.4930      & 0.1620          & -0.5787 \\
B1=0 vs Control & 4.83    & 0.1003       & 0.9225     &  0.0571     & 12.05 & -2.4799    & \textbf{0.0289}   & -1.1395 & 9.36     & -2.9059      & \textbf{0.0167} & -1.2651 \\
\toprule
\multicolumn{13}{c}{Non-Speaking Interval} \\
\midrule
                & \multicolumn{4}{c}{Gaze Engagement Ratio} & \multicolumn{4}{c}{Gaze Variance}    & \multicolumn{4}{c}{Gaze Concentration Area} \\
\midrule                    
                & df      & \textit{t}   & \textit{p}  &  \textit{d}   & df    & \textit{t} & \textit{p}  &  \textit{d}  & df       & \textit{t}   & \textit{p}   &  \textit{d}   \\
\midrule                    
B1=2 vs B1=0    & 34.27   & -1.8787      & 0.1041     &    -0.8268  & 41.00 & 3.5495     & \textbf{0.0009} & 0.6626 & 22.76    & 1.9160       & 0.0680   &  0.4360 \\
B1=2 vs Control & 27.23   & -2.9343      & \textbf{0.0094} & -0.8850 & 43.73 & 2.7611     & \textbf{0.0083} & 0.5423 & 14.08    & -2.1926      & \textbf{0.0456} & -0.7324 \\
B1=0 vs Control & 4.68    & -0.0364      & 0.9718      & -0.0206    & 12.88 & -1.3989    & 0.1854         & -0.6870  & 11.30    & -3.5962      & \textbf{0.0040} & -1.6238 \\
\bottomrule
\end{tabular}
\end{table*}

\begin{figure*}
 \centering
 \includegraphics[width = 1\textwidth]{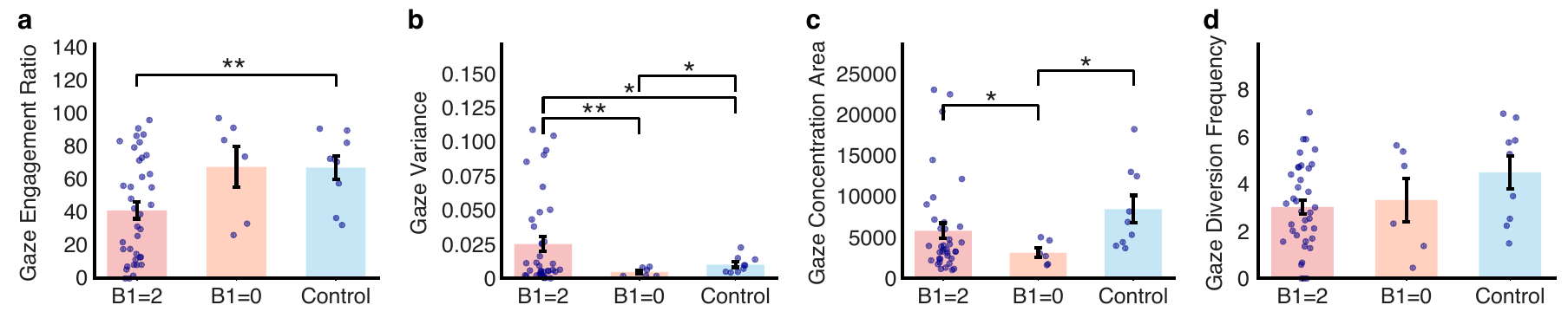}
 \caption{Summary of gaze metrics. (a) Gaze engagement ratio. (b) Gaze variance. (c) Gaze concentration area. (d) Gaze diversion frequency. Error bars denote ±SEM across samples. Each dot represents a sample. Asterisks indicate a significant difference using Welch's t-test. \small{*: $p < 0.05$, and **: $p < 0.01$}.}
 \label{gaze_feat} 
\end{figure*}

\begin{figure*}
 \centering
 \includegraphics[width = 1\textwidth]{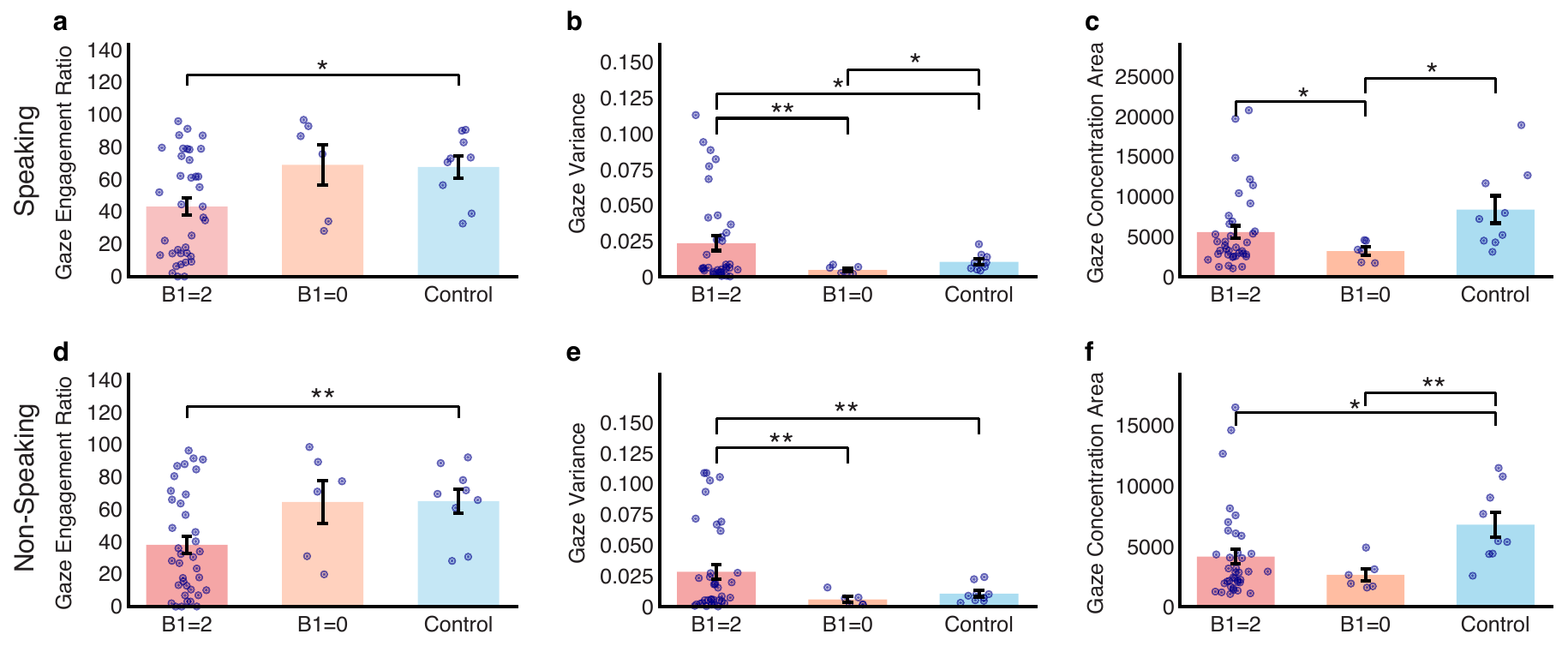}
 \caption{Summary of gaze metrics during speaking and non-speaking intervals. (a-c) Speaking. (d-f) Non-speaking. (a, d) Gaze engagement ratio. (b, e) Gaze variance. (c, f) Gaze concentration area. Error bars denote ±SEM across samples. Each dot represents a sample. Asterisks indicate a significant difference using Welch's t-test. \small{*: $p < 0.05$, and **: $p < 0.01$}.}
 \label{gaze_feat_speaking_not_speaking} 
\end{figure*}

\subsection{Gaze Engagement}
\label{gaze_engage}

The gaze engagement ratio is the proportion of time a participant's gaze focused on the examiner in relation to the total video duration, serving as a key measure of involvement and attentiveness during interactions (Figure \ref{gaze_feat}(a)). As shown in Table \ref{results_table}, the B1=2 group had an average gaze engagement ratio of 41.29\%±31.20\%. In contrast, the B1=0 group displayed a higher average of 67.67\%±30.44\%, while the control group showed an average ratio of 67.22\%±21.20\%. The control group had a significantly higher gaze engagement ratio than the B1=2 group, while there was no significant difference between the control and B1=0 groups (Figure \ref{gaze_feat}(a); see Table \ref{results_t-test} for statistics). These results indicate that a reduced proportion of gaze is a critical factor for atypical gaze in ASD.

We further analyzed the gaze engagement during speaking (Figure \ref{gaze_feat_speaking_not_speaking}(a)) and non-speaking (Figure \ref{gaze_feat_speaking_not_speaking}(d)) intervals. During both speaking and non-speaking intervals, the B1=2 group had a lower gaze engagement ratio (Table \ref{results_table_speaking_or_not}). Similar to the entire video, the control group had a significantly higher gaze engagement ratio than the B1=2 group for both speaking and non-speaking (Table \ref{t-test_speaking_not_speaking}) intervals, while there was no significant differences between the control and B1=0 groups (Figure \ref{gaze_feat_speaking_not_speaking}(a); Figure \ref{gaze_feat_speaking_not_speaking}(d)).

\subsection{Gaze Variance}
\label{gaze_var}

Gaze variance is a metric used to assess how steady participants' gazes were. It was calculated as the Euclidean distance between consecutive pairs of gaze angle coordinates (x, y), which represent the horizontal and vertical orientations of the participant's gaze in global coordinates, and then averaged across both eyes. This metric captures the fluctuation in the individual's gaze, offering insights into the dynamic aspects of attention and involvement during the interaction. The B1=2 group (0.0255±0.0333 [mean±SD]) had significantly greater variability compared to the B1=0 group (0.0048±0.0031; see Table \ref{results_t-test} for statistics; Figure \ref{gaze_feat}(b)) and compared to the control group (0.0103±0.0061; Table \ref{results_t-test}; Figure \ref{gaze_feat}(b)). Similar results were observed during both speaking and non-speaking intervals (Figure \ref{gaze_feat_speaking_not_speaking}(b); Figure \ref{gaze_feat_speaking_not_speaking}(e); see Table \ref{results_table_speaking_or_not} for values and Table \ref{t-test_speaking_not_speaking} for statistics). In contrast, the B1=0 group had significantly less variability compared the control group (Table \ref{results_t-test}; Figure \ref{gaze_feat}(b)), specifically during speaking intervals (Figure \ref{gaze_feat_speaking_not_speaking}(b)). However, there were no significant differences between the B1=0 and control groups during non-speaking intervals, indicating a convergence in gaze stability when not engaged in verbal communication. Together, our results suggest that increased gaze variability is an important factor in atypical gaze in ASD as identified on the ADOS-2.

\begin{figure*}
 \centering
 \includegraphics[width = 1\textwidth]{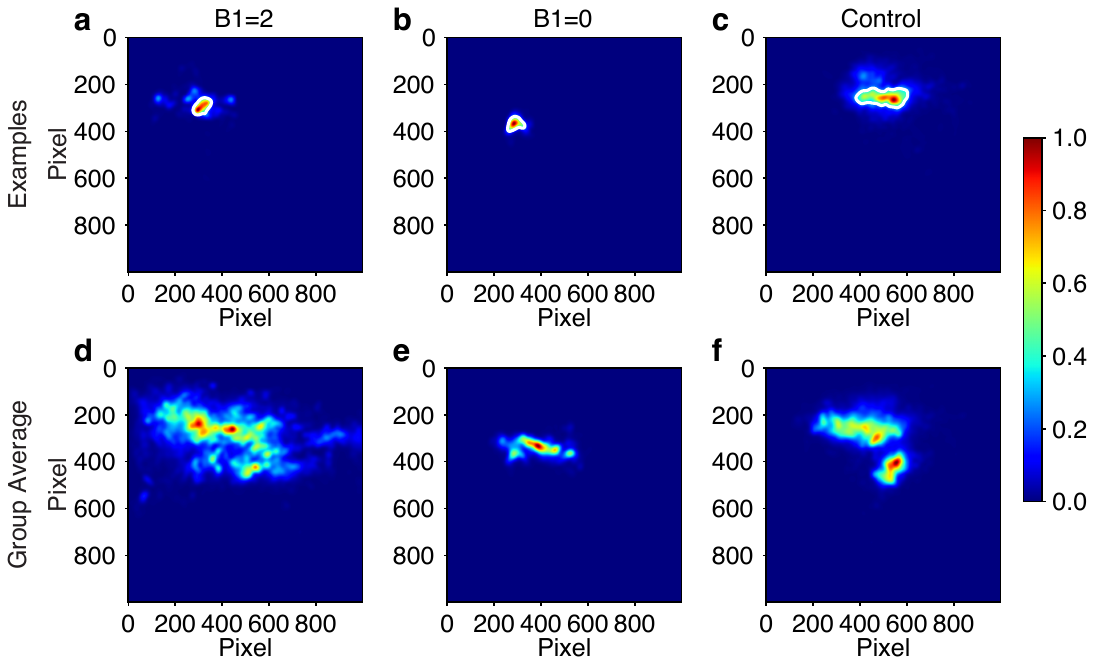}  
 \caption{Gaze density maps. (a-c) Example density maps from the three groups. (d-f) Average density maps from the three groups. (a, d) B1=2 group. (b, e) B1=0 group. (c, f) Control group. Each map shows the likelihood of looking at a given location. The scale bar (color bar) is in arbitrary units.}
 \label{density_example_density_avg_group} 
\end{figure*}

\begin{figure*}
 \centering
 \includegraphics[width = 1\textwidth]{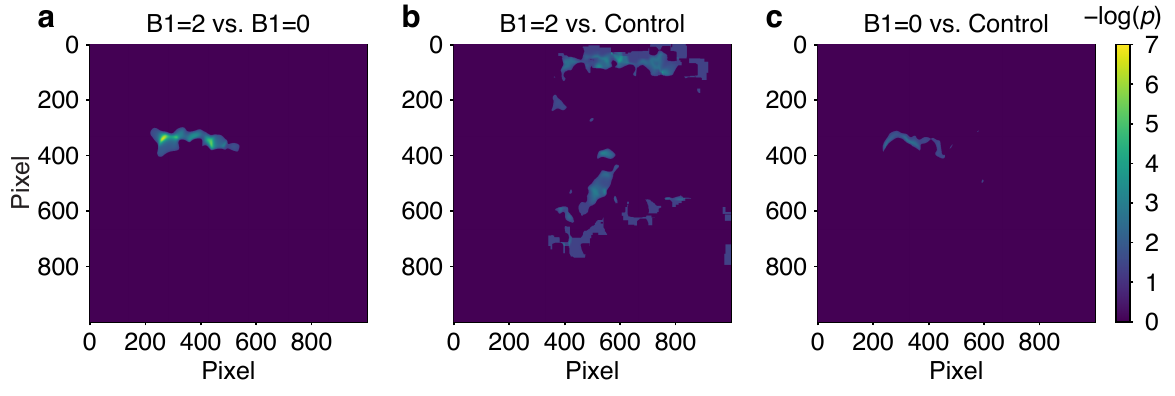} 
 \caption{Statistical maps. (a) B1=2 vs. B1=0. (b) B1=2 vs. Control. (c) B1=0 vs. Control. Statistical maps show areas that had significant difference in density maps between groups. Color coding shows the negative logarithm of p-values. Only significant pixels (\small{$p < 0.05$, uncorrected}) are shown.}
 \label{density_p_map} 
\end{figure*}

\subsection{Gaze Distribution and Gaze Concentration Area}
\label{gaze_area}

In this section, we conducted two analyses to examine gaze location. First, we examined distribution of gaze via qualitative comparisons of gaze density maps and statistically compared the distribution of gaze density pixel-by-pixel to reveal areas in the visual field that differed between groups. Second, we summarized the concentration (i.e., area/spread) of focused gaze in the gaze density map and compared these between groups.

We used gaze coordinates from the videos of each participant to construct individual gaze density maps. These maps visually represent the focal points of participants' gaze throughout the task, highlighting where their attention was predominantly directed. Figure \ref{density_example_density_avg_group}(a-c) show examples of these density maps, with one representative participant selected from each of the three groups. Following this individual analysis and normalization of the gaze data, we aggregated the results within each group to calculate an average density map. This process involved combining the gaze coordinates from all participants in a group and then generating a collective density map to represent the average gaze distribution. Figure \ref{density_example_density_avg_group}(d-f) illustrate these group-average density maps, providing a comparative visualization of gaze patterns across groups. To optimize visual clarity and detail, the gaze density maps were sized at 1000x1000 pixels, ensuring a detailed representation of gaze distribution patterns among participants. Qualitatively, we found that the B1=2 group had a wide distribution of gazes (Figure \ref{density_example_density_avg_group}), consistent with the results of gaze variability. Interestingly, the B1=0 group had the narrowest distribution of gazes (Figure \ref{density_example_density_avg_group}), also consistent with gaze variability. We further generated the statistical maps (i.e., the p-value maps) to compare gaze density across groups (Figure \ref{density_p_map}). The B1=0 group differed from both the B1=2 and control groups in similar areas of the visual field, while the B1=2 group differed from the control group more extensively in both the upper and lower visual fields.

To quantitatively assess the spread of gaze focus, we calculated the "gaze concentration area" as the pixel area of hot spots within the gaze density map (see Section \ref{methods}; Table \ref{results_table}). We observed an average gaze concentration area of 5849.16±5771.93 pixels for the B1=2 group, 3161.25±1442.79 pixels for the B1=0 group, and 8530.02±5077.82 pixels for the control group (Table \ref{results_table}). The B1=0 group had a significantly smaller gaze concentration area compared to both the B1=2 group (Figure \ref{gaze_feat}(c); see Table \ref{results_t-test} for statistics) and the control group (Figure \ref{gaze_feat}(c); Table \ref{results_t-test}). Similar results were observed during speaking intervals, but during non-speaking intervals, gaze concentration in the B1=2 group was more similar to the B1=0 group, and both groups had a significantly smaller gaze concentration area compared to controls (Figure \ref{gaze_feat_speaking_not_speaking}(c); Figure \ref{gaze_feat_speaking_not_speaking}(f); see Table \ref{results_table_speaking_or_not} for values and Table \ref{t-test_speaking_not_speaking} for statistics). Overall, our results revealed a consistently smaller gaze concentration area in the B1=0 group compared to the control group and a smaller gaze concentration area in the B1=2 group compared to the control group during non-speaking intervals. However, the B1=2 group did not differ significantly from the control group during speaking intervals. This pattern suggests that this feature of gaze does not fully explain examiner-perceived atypical gaze in ASD. 

\subsection{Gaze Diversion Frequency}
\label{gaze_freq}

Gaze diversion frequency was defined as how often individuals redirected their gaze away from the examiner's face per minute. As summarized in Table \ref{results_table}, the average gaze diversion frequency was 3.0385±1.8270 in the B1=2 group, 3.3338±2.2311 in the B1=0 group, and 4.5065±2.0894 in the control group (Figure \ref{gaze_feat}(d)). We observed no significant disparity in gaze diversion frequency between the B1=2 and B1=0 groups or between the B1=0 and control groups (Table \ref{results_t-test}; the comparison between the B1=2 and control groups approached marginal significance). Therefore, gaze diversion frequency could not explain atypical gaze in ASD.

\subsection{Classifier}
\label{classifier_sec}

\begin{table}[htbp] 
\centering
\resizebox{\columnwidth}{!}{
\begin{threeparttable}
\caption{Classification Using a Random Forest Algorithm.}
\label{classfier_}
\begin{tabular}{ccccc}
\toprule
\multicolumn{5}{c}{B1=2 vs. Controls} \\
\midrule
Group & Accuracy & Precision & Recall & F1 Score \\
\midrule
All\tnote{a} & 0.839 & 0.817 & 0.839 & 0.809 \\
B1=2 & 0.962 & 0.858 & 0.962 & 0.901 \\
Control & 0.379 & 0.510 & 0.353 & 0.387 \\
\midrule
Shuffled dataset (All)\tnote{b} & 0.756 & 0.646 & 0.756 & 0.690 \\
Shuffled dataset (B1=2) & 0.953 & 0.790 & 0.953 & 0.854 \\
Shuffled dataset (Control) & 0.000 & 0.000 & 0.000 & 0.000 \\
\toprule
\multicolumn{5}{c}{B1=2 vs. B1=0} \\
\midrule
Group & Accuracy & Precision & Recall & F1 Score \\
\midrule
All\tnote{a} & 0.831 & 0.747 & 0.831 & 0.781 \\
B1=2 & 0.970 & 0.852 & 0.970 & 0.903 \\
B1=0 & 0.041 & 0.055 & 0.033 & 0.039 \\
\midrule
Shuffled dataset (All)\tnote{b} & 0.827 & 0.749 & 0.827 & 0.782 \\
Shuffled dataset (B1=2) & 0.961 & 0.857 & 0.961 & 0.903 \\
Shuffled dataset (B1=0) & 0.000 & 0.000 & 0.000 & 0.000 \\
\toprule
\multicolumn{5}{c}{B1=0 vs. Controls} \\
\midrule
Group & Accuracy & Precision & Recall & F1 Score \\
\midrule
All\tnote{a} & 0.716 & 0.788 & 0.716 & 0.708 \\
B1=0 & 0.620 & 0.526 & 0.490 & 0.477 \\
Control & 0.842 & 0.736 & 0.800 & 0.720 \\
\midrule
Shuffled dataset (All)\tnote{b} & 0.336 & 0.327 & 0.336 & 0.297 \\
Shuffled dataset (B1=0) & 0.036 & 0.035 & 0.030 & 0.029 \\
Shuffled dataset (Control) & 0.610 & 0.445 & 0.580 & 0.451 \\
\bottomrule
\end{tabular}
\begin{tablenotes}
\item[a] Model performance: Shown are the performance of the random forest model trained with the full dataset ("All") and for each group.
\item[b] Model performance of the shuffling test.
\end{tablenotes}
\end{threeparttable}
}
\end{table}

We next evaluated the performance of a classifier trained on the quantified gaze features described above (see Section \ref{methods} for detailed methods). 

Fist, we compared between the B1=2 group and the control group (Table \ref{classfier_}). The overall accuracy of the observed model was 83.9\%, greater than the prediction accuracy of 75.6\% of the shuffling test. For the accuracy of individual groups, the classifier yielded 96.2\% for the B1=2 group and 37.9\% for the control group, compared to 95.3\% and 0\%, respectively, of the shuffling test. The high accuracy in the shuffling tests for the B1=2 group was because the model erroneously classifies all test samples as belonging to the B1=2 group (and thus 0\% for the control group). Although there was a bias in the classifier, the above-chance performance indicates that gaze features could distinguish individuals with ASD from controls.

Second, we compared between the B1=2 group and the B1=0 group (Table \ref{classfier_}). The accuracy of the observed model and the shuffled model were 83.1\% and 82.7\%, respectively, showing no successful learning. For each individual group, the observed model achieved an accuracy of 97.0\% for B1=2 and 4.1\% for B1=0, respectively, whereas the shuffled model achieved a similar accuracy of 96.1\% and 0\%, respectively. The high accuracy observed in the shuffled data might be due to the disproportionate sample sizes, with 37 in the B1=2 group compared to only 6 in the B1=0 group, leading to a model that predicted all labels as B1=2. Therefore, our classification model could not differentiate the B1=2 group vs. the B1=0 group.

Third, we compared between the B1=0 group and the control group (Table \ref{classfier_}). The overall prediction accuracy of the observed model was 71.6\%, substantially higher than the shuffled model (33.6\%). For both the B1=0 group (observed: 62.0\%; shuffled: 3.6\%) and control group (observed: 84.2\%; shuffled: 61.0\%), the accuracy of the observed model was higher than that of the shuffled model, suggesting successful learning. Therefore, even though individuals in the B1=0 group were considered to have normal gaze behavior in ADOS-2 assessments, there were still discernible differences compared to the gaze patterns of the control group, which could be detected by our gaze features and machine learning model.

\section{Discussion}

Naturalistic human behavior presents a complex, multidimensional signal that is invaluable for understanding social behavior in people with ASD. Video-based analyses offer valuable opportunities to study social functioning in ASD. In this study, we investigated the properties of social gaze in ASD participants using objective metrics and compared these metrics to controls. Through computer vision and machine learning, we quantitatively evaluated gaze features captured in third-person ADOS-2 interview videos of participants with ASD and controls. Specifically, we identified four measurable features related to eye contact for our analysis: gaze engagement, variance, area density, and diversion frequency. We found systematic differences in these objective gaze features across groups, underscoring their importance in ASD diagnosis. Using a mixed-method approach incorporating visual and auditory features into analyses, we demonstrated the practicality of using third-person videos to capture and analyze gaze behaviors in ASD.

\subsection{Result Interpretation}

The high detection ratio (Section \ref{detect_ratio}) validated the feasibility and reliability of using objective detection models to analyze gaze patterns in video content. Below, we discuss the implications of our results for each gaze metric, as well as for the machine learning model.

\subsubsection{Gaze Engagement}

As described in Section \ref{gaze_engage}, ASD participants with identified impairments in eye contact (B1=2 group) had significantly shorter social gaze durations compared to neurotypical individuals. Likewise, they had shorter gaze duration on average than the individuals with ASD who were not identified with impaired eye contact (B1=0 group). This suggests that this feature could effectively distinguish between neurotypical individuals and individuals with ASD who have impaired eye contact, serving as a vital indicator that is readily quantifiable. Moreover, the similarity in results between speaking and non-speaking intervals further validated the feasibility and reliability of using gaze duration to differentiate ASD individuals with impaired eye contact from non-ASD individuals. Additionally, as expected, within each group, gaze duration during speaking intervals was consistently higher than during non-speaking intervals.

\subsubsection{Gaze Variance}

As discussed in Section \ref{gaze_var}, gaze variance reflects the fluctuations and stability of a participant's gaze, and interestingly, the ASD subgroups diverged on this feature but both differed significantly from the neurotypical group. In comparison to the control group, gaze variance was greater in individuals with ASD and identified impairments in eye contact (B1=2 group). However, the B1=0 group, which had no diagnosis of eye contact impairment according to the ADOS-2 assessment, had significantly less gaze variance compared to neurotypical individuals and the B1=2 group. Similar patterns were observed during speaking intervals, aligning with the overall results. However, there were no significant differences between the B1=0 and control groups during non-speaking intervals, indicating a convergence in gaze stability when not engaged in verbal communication. This suggests that although unstable and overly-fixated gaze are both patterns of atypicality associated with ASD, ADOS-2 examiners may find it easier to recognize unstable gaze in participants than to identify subtle decreases in variability that are restricted to periods of verbal communication.

Notably, as shown in Figure \ref{gaze_feat}(b), some participants in the B1=2 group also had low gaze variance (i.e., similar to the B1=0 group), but individual examination of data from these participants confirmed that they also had a lower gaze engagement ratio compared to the B1=0 group. This suggests that given a similar gaze variance, individuals in the B1=2 group could still be differentiated by their different gaze engagement. This is also consistent with the idea that gaze metrics/features from different aspects should be used for ASD assessment. 

\subsubsection{Gaze Distribution and Gaze Concentration Area}

In Section \ref{gaze_area}, the gaze density maps indicate that participants from the group diagnosed with eye contact issues ($B1=2$) had a broader coverage area, suggesting their gaze was more dispersed. In contrast, the gaze of ASD participants without eye contact issues (B1=0) was the most focused, with a very limited range. These observations well align with the gaze variance results. While speaking and non-speaking conditions showed a similar result, there were also nuanced differences, highlighting the need for further analysis to understand gaze behavior in non-speaking contexts. In addition to differences in the extent of gaze distribution, the participant groups also exhibited variations in their preferred gaze locations, highlighting the importance of incorporating information about the content at these preferred gaze locations in future studies. Furthermore, future experiments will be needed to control for differences in experimental setup (e.g., camera distance, especially across sites).


\subsubsection{Gaze Diversion Frequency}

As shown in Section \ref{gaze_freq}, the differences in gaze diversion frequency among the three groups were not pronounced, with only a marginal difference observed between the B1=2 group and the control group. This suggests that this feature was relatively weak in distinguishing among the three groups. Additionally, the limitations posed by the third-person perspective of the videos may impede the accurate calculation of diversion frequency, further affecting the effectiveness of this feature in our analysis.

\subsubsection{ASD Classifier}

First, the comparison of predictive performance between the B1=2 group and the control group revealed a significant improvement in the predictions for the control group. The high accuracy in the shuffled results for the B1=2 group may stem from underfitting, where the model erroneously classified all test samples as belonging to the B1=2 group. The results demonstrated that gaze features effectively distinguished ASD individuals with atypical eye contact from neurotypical individuals, indicating the potential of these features in identifying autism-related gaze abnormalities. Second, we found it difficult to distinguish the B1=2 and B1=0 groups based on our classification model. Although individual gaze features such as gaze variance indicated a significant difference between the $B1=2$ and $B1=0$ groups, the selected gaze features were still insufficient for our machine learning model. Additionally, the high accuracy observed in the shuffled data was likely due to the bias of disproportionate sample sizes, with 37 in the B1=2 group compared to only 6 in the B1=0 group, leading to a model that predicted all labels as B1=2, thereby achieving high accuracy. Third, the results comparing the B1=0 group and the control group suggest that even though individuals in the B1=0 group were considered to have normal eye-contact behavior in ADOS-2 assessments, their gaze patterns differ from neurotypical individuals in manner that is detectable using our extracted gaze features.

\subsection{Advantages and Limitations}

This study combines the advantages of direct behavior observation in a naturalistic setting and objective machine-learning methodology. The ADOS is the gold standard of observational assessment of ASD. It is designed to capture abnormal behaviors common in people with ASD in a controlled but somewhat naturalistic setting and specifically aims to elicit eye contact during social interactions. By utilizing video recordings of standardized ADOS administration, we were able to apply machine learning to observed behavior.

Contrary to the predominant use of first-person perspective videos in ASD gaze research databases—wherein the examiner is often equipped with a head-mounted camera or the participant completes tasks directly in front of the camera with concurrent eye-tracking—we employed a third-person viewpoint. This approach can capture the interaction between the examiner and the participant within the same scene. In addition, third-person perspective videos are less costly to produce as they do not necessitate additional auxiliary equipment during recording, offering an economical alternative for extensive ASD research.

With the advantages of our study in mind, it is worth noting that the current study still has several limitations. The resolution of the video recordings was low and only from one perspective, which, despite achieving a high success rate in gaze detection, limits the predictions to basic directional coordinates of the participant's gaze. Higher video resolution from multiple cameras could enable the extraction of more detailed information, such as the examiner's gaze direction and facial expressions of both parties. Additionally, the current sample size was small (both in number of participants and amount of video footage), which may affect the reliability of feature analysis and classifier training. Longer video samples would enable further examination of gaze behavior during verbal communication, differentiating between when participant is speaking vs. listening to the examiner. Moreover, the grouping was solely based on the eye-contact-related scoring items from ADOS-2, which were used as labels for classifier training. Future studies could expand to not only predict whether a participant is likely to have ASD, but also identify variants of ASD based on subtle behavioral differences that may be overlooked in clinical assessment.

\subsection{Atypical Social Behaviors in ASD}

People with ASD exhibit atypical attention to social stimuli, such as faces, but prefer to gaze at inanimate objects \cite{wang2017social, wang2015atypical}. They also have unusual fixation patterns on faces, tending to look more towards the mouth while avoiding the eyes \cite{adolphs2013abnormal, kliemann2010atypical, klin2002visual, neumann2006looking, pelphrey2002visual, spezio2007abnormal, spezio2007analysis}. Additionally, preschool children with autism show a reduction in dyadic orienting and joint attention \cite{leekam2006dyadic}. More broadly, in addition to gaze, individuals with ASD demonstrate a range of atypical behaviors. First, facial expressions can be difficult for people with ASD to read and interpret \cite{cao2020flexible, webster2021posed, yu2022distinct}. In addition, people with ASD may have a distinct way of moving their facial muscles to express emotions, which can result in expressions that are not easily recognizable or may seem inconsistent with the emotions they are feeling \cite{gallagher2015conceptual, gallese2013mirror}. Second, people with ASD exhibit restricted and repetitive behaviors, such as hand and finger flicking \cite{leekam2011restricted}. They also display motor dysfunctions, including postural instabilities, atypical gait, mistiming of motor sequences, problems with motor coordination, difficulties with anticipatory postural adjustments, and expressionless faces \cite{gallese2013mirror}. Third, people with ASD may exhibit atypical speech patterns, including a monotonic tone \cite{kissine2021phonetic}, an unusual usage of words \cite{li2019automated}, and in some cases, a louder voice \cite{brownell2002musically}. Future computer vision and machine learning studies are needed to investigate other atypical social behaviors in individuals with ASD.

\subsection{Computer Vision Analysis of ASD Behavior}

Recently, computer vision and machine learning have been employed to analyze various behavioral markers in ASD. For example, deep learning can detect hand flapping from unstructured home videos to aid in diagnosis \cite{lakkapragada2022classification}. Computer vision analysis has revealed that autistic toddlers exhibit differences in their head movement dynamics when viewing audiovisual stimuli \cite{krishnappa2023complexity}. Furthermore, computer vision analysis has been applied to assess motor imitation and imitative learning \cite{lidstone2021automated, tunccgencc2021computerized, zampella2021computational} and capture atypical attention in toddlers with autism \cite{campbell2019computer}. Natural language processing on electronic health records has also been used to develop a phenotype ontology for ASD \cite{zhao2022development}. Lastly, machine learning can characterize the classification boundary between ASD and non-ASD \cite{tuncc2021diagnostic}, and a tele-assessment for ASD in response to the disruptions caused by the COVID-19 pandemic, TELE-ASD-PEDS \cite{wagner2021use}, was developed through the application of machine learning to a large clinical database of several hundred children \cite{adiani2019usability}.

\subsection{Future Work}

This study successfully utilized an objective measurement model to extract, process, and analyze gaze features from a third-person perspective in ADOS-2 interview videos. The model was trained to classify ASD participants with gaze issues from neurotypical participants accurately.Additionally, the findings regarding the speaking duration ratio showed that neurotypical participants typically had longer speaking times than participants with ASD. This suggests that neurotypical participants may be more conversational, which could be substantiated by further research. In future work, there is potential to expand the analysis to incorporate other features, such as facial expressions, body movements, and more detailed speech transcription, and to consider the development of multimodal classification models. For example, idiosyncratic interests observed in ASD \cite{sasson2016brief, sasson2014visual, sasson2008children} can be reflected in both atypical gaze and action towards those objects (it may be reflected in speech as well). A multimodal data fusion technique can be developed to comprehensively and effectively analyze the deficits from various aspects. Additionally, this study demonstrated the utility of third-person perspective videos of ASD participants in gaze analysis. These videos are relatively low-cost to obtain, more universally applicable, and have the potential to be extended to home video-based ASD detection and analysis.

\section{Methods}
\label{methods}
\subsection{Datasets}

In this study, we utilized two video datasets: the Caltech ADOS-2 Video Dataset and the West Virginia University (WVU) ADOS-2 Video Dataset. 
The Caltech dataset contained ADOS-2 interviews with 30 high-functioning adults (24 males) diagnosed with ASD. The WVU dataset contained videos from a 9 neurotypical control participants (3 males) who had no psychiatric disorders and served as the control group for the ADOS-2 interview recordings. 
All participants gave their informed consent by the Institutional Review Board (IRB) protocols of Caltech and WVU. 

In the Caltech dataset, 30 high-functioning participants with ASD completed ADOS-2 interviews. Seven participants contributed two samples and three participants contributed three samples, leading to a total of 43 samples. All participants met the Diagnostic and Statistical Manual of Mental Disorders, Fifth Edition (DSM-V) / International Classification of Diseases, Tenth Revision (ICD-10) diagnostic criteria for ASD, and met the cutoff scores for ASD on the ADOS-2 revised scoring system for Module 4 \cite{hus2014autism}. ADOS-2 was scored according to the latest algorithm, and we also derived severity scores (social affect [SA]: 8.28±4.49 [mean±SD], restricted and repetitive behavior [RRB]: 2.44±1.48, severity score for social affect [CSS-SA]: 6.0±2.49; severity score for restricted and repetitive behavior [CSS-RRB]: 5.98±2.38, severity score for social affect plus restricted and repetitive behavior [CSS-All]: 5.65±2.76). The ASD group had a full-scale IQ (FSIQ) of 96.83±13.48 (from the Wechsler Abbreviated Scale of Intelligence-2), a mean age of 23.45±4.76 years, a mean Autism Spectrum Quotient (AQ) of 23.18±7.11, and a mean SRS-2 Adult Self Report (SRS-A-SR) of 86.23±24.03.

\subsection{Experimental Procedure}

Autism Diagnostic Observation Schedule, 2nd Edition (ADOS-2) is a widely-used gold standard for diagnosing ASD in clinical practice, which involves professionals who evaluate the participant's specific social and communicative behaviors to identify characteristics associated with ASD. ADOS-2 consists of structured and semi-structured interactions conducted by trained examiners who ensure the accuracy and consistency of evaluations. The examiners carefully observe the participant's responses and behaviors throughout the assessment. Based solely on observations during the assessment, examiners score the participant on items tailored to capture behaviors associated with the core symptoms of ASD, such as social interaction, communication skills, and restricted and repetitive behaviors. 

All ADOS-2 assessments used in this study were conducted by research-reliable administrators in a research setting using standardized administration of Module 4. ADOS-2 Module 4 is intended for participants with fluent speech, including adolescents and adults. It consists of 15 semi-structured interactions, all of which were administered and included in the videos. 

The ADOS-2 assessments were conducted in a controlled environment following in the ADOS-2 standards, within a quiet and private room. Only one examiner and one participant were present, seated face-to-face at a table, promoting a one-on-one interaction between the examiner and the participant. A video camera was placed nearby to clearly record the participant's physical behaviors, facial expressions, gestures, eye contact/gaze behavior, speech patterns, and reciprocal social interactions with the examiner. To minimize potential distraction, the camera was placed slightly outside the participant's direct line-of-sight

For the Caltech dataset, the camera was set to record at a data rate of 9.1 Mbps and a frame rate of 30 frames per second, with a video resolution of 720 × 480. All videos were captured at the California Institute of Technology. For the WVU dataset, the camera had a resolution of 1920 x 1080 and a frame rate of 60 frames per second. All videos were captured at the West Virginia University. The two sites had similar settings for video recordings.

\subsection{Assessment of ADOS Videos}

Based on all observations from the entire assessment, examiners provided scores for all ADOS-2 items immediately following each administration and generated summary scores using the current diagnostic algorithm \cite{hus2014autism}. Scored items capture specific aspects of behavior that fit within 5 broad categories: (A) language and communication, (B) reciprocal social interaction, (C) Imagination, (D) Stereotyped behaviors and restricted interests, and (E) other abnormal behaviors. However, the diagnostic algorithm utilizes items from all of these categories and groups them into two domains: Social Affect and Restricted Repetitive Behaviors. Item ratings most closely index specific behaviors and therefore serve as the ground truth for our machine learning.

Specifically, our study was centered on the eye contact behavior of individuals with ASD. Thus, we utilized the scores on the item that addresses gaze behavior, the Unusual Eye Contact item (B1 within the  Reciprocal Social Interaction category). This particular item distinguishes between a gaze that is clear, flexible, and socially modulated, employed for a variety of reasons, and a gaze that shows limitations in flexibility, suitability, or contexts. This item can be scored either a 0 or a 2. A score of 0 implies that the participant exhibits suitable gaze with minor alterations incorporated within other communication elements, whereas a score of 2 indicates that the participant employs inadequately modulated eye contact to start, end, or control social interaction. Among the 43 samples in the Caltech dataset, 37 samples had a B1 score of 2 and 6 samples had a B1 score of 0.


\subsubsection{Data}

This study analyzed eye contact during Task 6 of the ADOS-2 Module 4 (i.e., "Social Difficulties and Annoyance"). During this task, the participant is asked a range of questions related to social challenges and annoyances, such as interpersonal difficulties at home or school, instances of being bullied or teased, and how the participant reacts to these situations. The aim of the task is to assess the participant's understanding of social obstacles and their strategies to cope or react to them. This task was selected because it is the first instance of a purely question-and-answer format in the ADOS-2 interview sequence. Thus, the format encourages eye-to-eye contact and its placement in the assessment minimizes potential of tester fatigue. The Caltech dataset contained 43 videos from Task 6, which summed up to 267 minutes and 481,005 frames. The WVU dataset, on the other hand, had a total of 9 videos that summed up to 84 minutes and 302,055 frames.

\subsection{Data Preprocessing}

The original videos comprised both visual and audio components, which we separated using the function "ffmpeg" to isolate the visual and audio tracks. The visual frames were processed as continuous segments for subsequent analysis, allowing for the examination of spatio-temporal features such as head tracking.

FFmpeg is a free and open-source software project consisting of a vast software suite of libraries and programs for handling video, audio, and other multimedia files and streams. At its core, FFmpeg is designed to convert multimedia files between formats, resize videos, stream audio and video, and capture and encode real-time video and audio. We utilized ffmpeg to preprocess the original source clips, specifically to separate the visual and audio components while maintaining the original quality. Additionally, some videos were rescaled to two to three times of their original resolution to enhance the detection accuracy of subsequent analyses using models such as OpenFace and OpenPose.

\subsection{Feature Extraction}

To derive raw features, we fed video clips that included only the participant into OpenFace, which allowed us to pull out gaze coordinate data for each frame. However, as the examiner was frequently positioned sideways or with their back facing the camera, OpenFace did not provide reliable results. Consequently, we utilized OpenPose to handle the examiner’s videos, enabling us to secure the head position coordinates, which were pivotal for feature computations. For the audio data, we fed the recordings into Google's speech-to-text service to secure transcriptions of the interactions between the examiner and the participant. This transcription aided us in precisely identifying the video segments where speaking took place. Specifically,

\subsubsection{OpenFace}
\label{detect_ratio}

OpenFace 2.0 \cite{baltrusaitis2018openface} is an open-source, high-performance facial behavior analysis tool, celebrated for its state-of-the-art approach in automatic facial landmark detection, head pose estimation, and advanced real-time gaze tracking. By applying OpenFace, we discernibly increased the efficiency and quality of feature extraction, reliably gathering topographical facial data and information for algorithmic media behavior. Using the OpenFace toolkit, we extracted features related to the participant's gaze, including the gaze direction coordinates and the eye position coordinates.

As shown in Table \ref{table_ratio}, the success ratio of gaze detection, which is determined by the percentage of frames where participants' gaze was accurately detected by OpenFace out of the total number of frames, was high for all three groups. The B1=2 group achieved a detection ratio of 95.38\%±7.82\%, while the B1=0 group demonstrated a higher accuracy at 99.64\%±0.55\% (see Table \ref{t-test_speaking_ratio} for statistics; similar results were derived when matching the detection ratio). The control group also exhibited consistent performance, with a detection ratio of 98.44\%±2.81\%. When all groups were combined, an overall success ratio of 96.40\%±6.87\% was achieved, indicating reliable gaze detection across different videos and datasets.

\subsubsection{OpenPose}

OpenPose \cite{cao2017realtime} represents a pioneering real-time system designed for multi-person 2D pose estimation, extending its capabilities to detect body, face, and hand keypoints from single images or video streams. This open-source tool leverages deep learning algorithms to accurately identify and track human body parts, enabling advanced studies in computer vision, human-computer interaction, and motion analysis. Due to the limitations of the third-person perspective, the examiner in the videos was often shown in profile or with their back facing the camera, making it challenging for OpenFace to capture facial and eye features accurately. Therefore, we utilized OpenPose to extract the coordinates of head-related keypoints.

\subsubsection{Google Speech-to-Text}

Google Speech-to-Text \cite{GoogleCloud2021} is a cutting-edge cloud-based service that leverages powerful deep learning neural network algorithms to convert audio to text. Developed by Google, this tool is designed to provide high accuracy in transcribing spoken words into written text across a wide array of languages and dialects. We employed Google Speech-to-Text to process the audio segments of our samples, converting spoken interactions into text while also recording the corresponding timestamps. This allowed us to later integrate audio information for a comprehensive analysis of the relationship between participant gaze behaviors and their conversational dynamics.

As demonstrated in Table \ref{table_ratio}, the speaking duration ratio was defined as the portion of time that the examiner and participant were engaged in speech (i.e., either the examiner or the participant was speaking) throughout the video. B1=2 group exhibited a mean speaking duration ratio of 0.6419±0.1508, B1=0 group showed a mean of 0.6834±0.0651, and the control group demonstrated a mean of 0.8300±0.0629. When combining all groups, the overall mean was determined to be 0.6792±0.1487. Subsequent Welch's t-tests were conducted to compare the speaking duration ratios between groups (Table \ref{t-test_speaking_ratio}). The comparison between the B1=2 and B1=0 groups indicated no significant difference in speaking duration ratios (\( t(15.81) = -1.1405, \, p = 0.2711 \)). However, significant differences were observed when comparing the B1=2 group with the control group (\( t(32.07) = -5.7911, \, p = 1.9736\times10^{-6} \)) and B1=0 with the control group (\( t(10.59) = -4.3289, \, p = 0.0013 \)), suggesting that the control group's speaking duration ratio was significantly higher than the ASD groups. 

\begin{table}[]
\caption{Mean and standard deviation of the gaze detection ratio and speaking duration ratio across the three groups.}
\label{table_ratio}
\centering
\begin{tabular}{ccc}
\toprule
        & Gaze Detection Ratio & Speaking Duration Ratio \\
\midrule         
B1=2    & 95.38\%±7.82\%     & 0.6419±0.1508         \\
B1=0    & 99.64\%±0.55\%     & 0.6834±0.0651         \\
Control & 98.44\%±2.81\%     & 0.8300±0.0629         \\
All     & 96.40\%±6.87\%     & 0.6792±0.1487         \\
\bottomrule
\end{tabular}
\end{table}

\begin{table}[]
\caption{Pairwise comparisons of the gaze detection ratio and speaking duration ratio across the three groups using Welch's t-test. ``df'' represents degrees of freedom, ``t'' is the t-statistic, ``p'' signifies the p-value, and ``d'' denotes the Cohen's d value. \textbf{Bold} indicates $p < 0.05$.}
\label{t-test_speaking_ratio}
\resizebox{\columnwidth}{!}{
\centering
\begin{tabular}{ccccccccc}
\toprule
                & \multicolumn{4}{c}{Gaze Detection Ratio} & \multicolumn{4}{c}{Speaking Duration Ratio} \\
\midrule             
                & \textit{df}          & \textit{t}            & \textit{p}    &  \textit{d}     & \textit{df}           & \textit{t}             & \textit{p}      &   \textit{d}   \\
\midrule             
B1=2 vs B1=0    & 37.98       & -3.2633      & \textbf{0.0023}  &  -0.5810  & 15.81        & -1.1405       & 0.2711     &  -0.2897 \\
B1=2 vs Control & 37.14       & -1.9220      & 0.0622   &  -0.4264 & 32.07        & -5.7911       &     \textbf{1.9736}$\mathbf{\times 10^{-6}}$ & -1.3526 \\
B1=0 vs Control & 8.90        & 1.2433       & 0.2455   & 0.5369  & 10.59        & -4.3289       & \textbf{0.0013}    & -2.2986  \\
\bottomrule
\end{tabular}
}
\end{table}

\subsection{Gaze Metrics}

Inspired by conventional autism assessments concerning eye contact, such as the duration of eye contact (either reduced or excessive), scope, and consistency, we designed several metrics for quantitative analyses using raw gaze features (see below). By examining these metrics, we could contrast the gaze behaviors between individuals with ASD and controls. We also implemented a machine learning model using these metrics to predict ADOS-2 scores associated with atypical gaze in ASD. It is worth noting that these metrics were solely obtained through computer algorithms applied to videos, which can eventually provide a cost-efficient method to improve diagnoses.

Given that people with ASD may display distinct behaviors while talking and being silent, such as challenges in sustaining eye contact while talking and possibly avoiding eye contact when not speaking, we explored the eye movements of the participants both during and outside the conversational phases. We first computed the portion of time that the examiner and participant were engaged in speech in the video to ensure an adequate period for interactive evaluation and to delve deeper into gaze behaviors during speaking (i.e., either the experimenter or the participant was speaking) and non-speaking (i.e., neither the experimenter nor the participant was speaking) segments.

\subsubsection{Gaze Engagement Ratio}

Many research studies have shown that people with autism tend to make less eye contact compared to neurotypical individuals \cite{yu2024multimodal, Wang2017}. Therefore, using videos captured from a third-person viewpoint, where the participant was fully visible while the examiner was seen in profile, we calculated the proportion of time that the participant gazed at the examiner in relation to the total video length.

We further investigated gaze engagement when someone was speaking and when no one was speaking, comparing these behaviors in ASD with controls. Our goal was to understand how the patterns of eye contact and gaze direction changed when moving from verbal to non-verbal communication in individuals with ASD. While speaking, participants may exhibit unique gaze tendencies, such as reduced or irregular eye contact, in contrast to the consistent and situationally appropriate gaze interactions seen in neurotypical speakers. On the other hand, in non-speaking situations, individuals with ASD may demonstrate different gaze behaviors, possibly looking away from the examiner or, at the other extreme, staring, which deviates from the socially connected gaze typically expected in neurotypical individuals. This analysis could highlight subtle variations and atypical patterns of gaze engagement used by individuals with ASD and demonstrate how these strategies differ between speaking and non-speaking contexts.

\subsubsection{Gaze Variance}

Gaze variance refers to the variability of where a person is gazing. This parameter, which is computed as the Euclidean distance between consecutive (x, y) coordinates in different frames, indicates how steady or fluctuating an individual's gaze is in space. Gaze variance provides insights into the visual tracking and attention concentration behaviors of individuals with ASD. Similar to the approach taken in studying the gaze engagement ratio, we also investigated gaze variance during speaking and non-speaking contexts.

\subsubsection{Gaze Distribution and Gaze Concentration Area}

Whereas gaze variance looks at \textit{movement} of gaze, the gaze density metric documents where a person's gaze \textit{rests} during specific epochs of the assessment. Gaze density was calculated by generating a density map (heat map) based on the x and y coordinates of the gaze over frames to define the gaze area. We then outlined the borders of the area where the gaze was concentrated and measured its size to assess the intensity of the gaze. This measurement quantified the distribution of gaze focus in space and the degree and focus of this distribution. We also examined gaze density during speaking and non-speaking intervals to explore differences in the visual attention patterns of participants across these two scenarios.

To generate statistical maps, for each pixel in the density maps, we had a distribution of values derived from each group. We conducted pair-wise t-tests for each pixel, comparing between two groups at a time, which yielded a p-value for each pixel. This procedure was repeated for each pair of groups, resulting in three p-value maps corresponding to the pairwise comparisons. Each map, maintaining a resolution of 1000x1000 pixels, underwent a transformation using the negative logarithm of the p-value to enhance visual interpretability. 

We defined gaze concentration area as the regions where the density surpassed 30\% of the maximum value observed in the gaze density map, thus emphasizing areas with notably high gaze concentration and offering a quantitative evaluation of focused gaze behavior. Note that gaze concentration area was calculated for each individual participant.

\subsubsection{Gaze Diversion Frequency}

Gaze diversion frequency was defined as the rate at which the participant shifted their gaze away from the face of the examiner within a minute. Unlike other features, since speaking and non-speaking intervals alternated frequently, we could not accurately calculate the gaze diversion frequency exclusively within these segments. Therefore, we did not incorporate audio components into the investigation of gaze diversion frequency.

\subsection{Classification}

We constructed a classifier using combined gaze features to differentiate groups. Given the dimensionality of the features and their hand-crafted nature, we selected the random forest algorithm as our classification model. The random forest classifier operates by constructing multiple decision trees during training and outputting the class that is the mode of the classes of the individual trees, making it particularly suitable for our experiment. We framed the task as a binary classification problem, training the classifier pairwise across the three groups. For each pair, all samples were randomly divided into disjoint training (80\%) and testing (20\%) sets. The features were normalized before training the random forest classifier to evaluate its performance. This process was repeated 100 times, with the average results being reported. As a benchmark, we also conducted a comparison where all sample labels were shuffled, followed by the same training and testing procedure. This allowed us to assess any performance improvement and validate the predictive relevance of the gaze features in distinguishing between the groups. 

We utilized four key metrics to show the performance of the model: Accuracy, which measures the proportion of correctly predicted observations to the total observations, providing an overall success rate of the classifier. Precision, defined as the ratio of correctly predicted positive observations to the total predicted positives, assesses the classifier's ability to identify only relevant instances. Recall, also known as sensitivity, indicates the ratio of correctly predicted positive observations to all actual positives, evaluating the classifier's capacity to find all relevant cases. Lastly, the F1 Score is the harmonic mean of precision and recall, offering a balance between the two, especially useful when the class distribution is uneven.

\subsection{Statistics}

Considering the diverse sample sizes and variances among the three datasets, we chose to use Welch's t-test to compare the means between two groups, as it compensates for these differences and is resilient against the presumption of equal variances, thereby determining if significant differences exist.

Although some participants contributed more than one sample and we treated each sample as independent in our analysis, we derived similar results when using unique participants.

\section*{Acknowledgments}
We thank Oana Tudusciuc, Laura Harrison, and Damian Stanley for administering the ADOS-2 interview. This research was supported by the NSF (BCS-2401398, IIS-2401748) and NIH (R01MH129426). The funders had no role in study design, data collection and analysis, decision to publish, or preparation of the manuscript.

\section*{Author Contributions}
X.Y., L.K.P., X.L., and S.W. designed research. L.K.P. performed experiments. X.Y., M.R., C.H., W.L., X.L., and S.W. analyzed data. X.Y., X.L., and S.W. wrote the paper. All authors discussed the results and contributed toward the manuscript.

\section*{Competing Interests Statement}
The authors declare no conflict of interest.

\bibliographystyle{IEEEtran}
\bibliography{0_main}{}

\newpage

\vfill

\end{document}